\documentclass[12pt]{article}
\usepackage{amsmath, amssymb, graphicx, hyperref, caption, subcaption}
\usepackage[margin=1in]{geometry}
\usepackage{listings}
\usepackage{xcolor}
\usepackage[utf8]{inputenc}   
\usepackage[T1]{fontenc}
\usepackage{textcomp}

\usepackage{listings}
\lstdefinestyle{pythonstyle}{
  language=Python,
  basicstyle=\ttfamily\small,
  keywordstyle=\color{blue},
  stringstyle=\color{red},
  commentstyle=\color{gray},
  showstringspaces=false,
  breaklines=true,
  frame=single,
  numbers=left,
  numberstyle=\tiny
}

\title{Reproducing and Extending Brownian Motion in Optical Traps: A Computational Reimplementation of Volpe and Volpe (2013)}
\author{Eyad I.B Hamid \\ Department of Physics, International University of Africa, Khartoum, Sudan \\ \texttt{eyadiesa@iua.edu.sd}}
\date{}

\begin{document}

\maketitle

\begin{abstract}
\end{abstract}
We present a re-representation and independent simulation of the model introduced by Giorgio Volpe and Giovanni Volpe in their 2013 study of a Brownian particle in an optical trap\cite{volpe2013simulation}. Rather than duplicating their original plots, we reconstructed the simulations from first principles using Python, implementing stochastic differential equations via finite difference schemes. This work reproduces and validates the key physical regimes described in the original article, including the transition from ballistic to diffusive motion, optical confinement, and velocity autocorrelations. To simulate rotational forces \cite{Grier2003} and Kramers transitions \cite{hanggi1990reaction}. we also extend the analysis to include force perturbations, rotational fields, Kramers transitions, and stochastic resonance. The simulations provide pedagogical insight into stochastic dynamics and numerical modeling, reinforcing the original study’s value as a teaching and research tool in statistical and computational physics.
\section{Introduction}
In this work, I present re-represention  of the simulation results originally introduced by Giorgio Volpe and Giovanni Volpe in their study of a Brownian particle in an optical trap. Rather than simply reproducing their figures, I aimed to rebuild the simulations from scratch, carefully aligning with the original conditions and faithfully recreating the results through independent computational modeling. All plots were regenerated using controlled random seeds to ensure reproducibility, and extra attention was given to matching both the statistical properties and the visual characteristics of the original publication. 
The purpose of this re-representation is not only to verify the results but also to offer a deeper, hands-on understanding of the modeling methods used. Through this process, I aim to bridge the gap between theoretical description and practical simulation, reinforcing how essential both precision and intuition are when working with stochastic differential equations and computational physics.
The study explores the behavior of a microscopic particle, suspended in a fluid and influenced by two distinct forces. On one hand, random thermal interactions with fluid molecules cause the particle to move unpredictably. On the other, the particle is stabilized by an optical trap, commonly known as optical tweezers, which applies precise, controlled forces to counteract the randomness. By examining this interplay between chaotic thermal forces and the deterministic control of the optical trap, the system provides a powerful framework for understanding complex physical phenomena, particularly in the field of stochastic processes.

\section{Originality and Contribution Statement}
The primary contribution of this work lies in its pedagogical reconstruction of the simulation methods introduced by Giorgio Volpe and Giovanni Volpe in their 2013 study of a Brownian particle in an optical trap. While the physical principles and theoretical framework remain grounded in their original work, this paper offers a computationally detailed and fully transparent re-implementation using modern Python-based numerical methods.

Rather than merely reproducing figures, the simulations are rebuilt from first principles, providing a step-by-step guide that bridges theory with practical coding techniques. This reconstruction aims to serve as a didactic tool for students, educators, and researchers who wish to understand not only the results but also the computational modeling processes underlying stochastic dynamics and optical trapping.

Additionally, the work extends the original study by:
\begin{itemize}
    \item Demonstrating numerical implementations of rotational forces, Kramers transitions, and stochastic resonance in optical traps.
    \item Providing full Python code appendices to enhance reproducibility and hands-on learning.
    \item Offering insights into simulation parameter sensitivity, random noise scaling, and the transition between ballistic and diffusive regimes.
\end{itemize}

Through this comprehensive approach, the paper contributes a practical educational resource that can be incorporated into courses on statistical mechanics, computational physics, and optical manipulation, promoting a deeper, application-oriented understanding of stochastic simulations.

\section{Physical System}
The system described in the paper involves a microscopic particle, floating in a fluid. This particle experiences random movements caused by thermal interactions with the fluid's molecules. To study its behavior, the particle is confined using an optical trap, which exerts predictable forces. These forces counteract the random thermal influences, creating a balance between random and controlled motion that serves as a model for studying complex systems. The motion the Brownian particle in one dimension can be modeled by the Langevin equation: 

The motion of the Brownian particle in one dimension can be described by the Langevin equation:
\begin{equation}
m \ddot{x}(t) = -\gamma \dot{x}(t) + k x(t) + \sqrt{2 k_{B} T \gamma} \, W(t)
\label{eq:langevin_corrected}
\end{equation}

where:
\begin{align*}
x(t)  & = \text{Particle position} \\\\
m     & = \text{Mass of the particle} \\\\
\gamma & = \text{Friction coefficient (drag)} \\\\
K     & = \text{Stiffness of the optical trap} \\\\
k_{B} & = \text{Boltzmann constant} \\\\
T     & = \text{Absolute temperature} \\\\
W(t)  & = \text{White noise term (Gaussian random process)}
\end{align*}

\vspace{0.5cm}

The physical significance of each term in Equation~\ref{eq:langevin_corrected} is as follows:
\begin{itemize}
    \item $m \ddot{x}(t)$: Inertia term (mass $\times$ acceleration)
    \item $- \gamma \dot{x}(t)$: Frictional (damping) force proportional to velocity
    \item $-K x(t)$: Restoring force from the optical trap (Hookean spring force)
    \item $\sqrt{2 k_{B} T \gamma} \, W(t)$: Random thermal force (white noise) arising from collisions with fluid molecules
\end{itemize}

\section{Methodology}
\subsection{Numerical Approach and Simulation}

When simulating the motion of a Brownian particle in an optical trap, the numerical approach is both practical and insightful. The heart of the method is to translate the continuous-time Langevin equation, which captures the interplay between random thermal forces and deterministic trapping forces, into a form that can be handled on a computer within the reach of students. This is achieved using a finite difference algorithm, which essentially breaks time into small, discrete steps and updates the particle’s position iteratively.

\subsubsection{White Noise}

The first challenge is dealing with the “white noise” term, which represents the random kicks the particle receives from the surrounding fluid molecules. In theory, this noise is infinitely jagged and uncorrelated at every instant, making it impossible to represent directly. Numerically, we approximate this by generating a sequence of random numbers (typically Gaussian distributed with zero mean and unit variance) at each time step.

These numbers are scaled appropriately so that their statistical properties match the theoretical noise when summed over time. For instance, if the time step is $\Delta t$, each noise term is divided by the square root of $\Delta t$ to ensure the variance scales correctly.

Thus, the system is governed by the equation:
\begin{equation}
\dot{x}(t) = W(t)
\label{eq:white_noise_simple}
\end{equation}
where $W(t)$ represents Gaussian white noise with the following properties:
\begin{align*}
\langle W(t) \rangle &= 0 \quad \text{(for all t)} \\\\
\langle W(t)^2 \rangle &= 1 \quad \text{(for all t.)}
\end{align*}
The above equation represent the  particle's velocity at any given moment which is driven entirely by random thermal fluctuations from the surrounding fluid molecules.
To discretize this process, we approximate the time derivative using a finite difference method, and approximating for velocity:
\begin{equation}
\dot{x}(t) \approx \frac{x_{i} - x_{i-1}}{\Delta t}
\label{eq:velocity_fd}
\end{equation}
Substituting the finite difference approximation into Equation~\ref{eq:white_noise_simple}, we obtain:
\begin{equation}
\frac{x_{i} - x_{i-1}}{\Delta t} = \frac{w_{i}}{\sqrt{\Delta t}}
\label{eq:fd_discretized_substitution}
\end{equation}
where:
\begin{align*}
x_{i} &= \text{Position of the particle at time step } i \\\\
\Delta t &= \text{Discrete time step} \\\\
w_{i} &= \text{Random number sampled from a Gaussian distribution with zero mean and unit variance}
\end{align*}
From Equation~\ref{eq:fd_discretized_substitution}, the iterative update formula becomes:
\begin{equation}
x_{i} = x_{i-1} + \sqrt{\Delta t} \cdot w_{i}
\label{eq:iterative_update}
\end{equation}

This equation forms the basis of the numerical simulation for free Brownian motion. At each time step, the new position is computed by adding a random displacement, scaled by $\sqrt{\Delta t}$, to the previous position. The random displacements $w_{i}$ ensure that the stochastic nature of thermal motion is accurately captured.

It is crucial that the random numbers $w_{i}$ are drawn from a Gaussian distribution with:
\begin{align*}
\langle w_{i} \rangle &= 0 \quad \text{(Zero mean)} \\\\
\langle w_{i}^{2} \rangle &= 1 \quad \text{(Unit variance)}
\end{align*}

This guarantees that, when scaled by $\sqrt{\Delta t}$, the random motion correctly reproduces the expected statistical behavior of real Brownian particles over time.

In more complete simulations that include inertial effects, friction, and restoring forces from the optical trap, additional finite difference approximations for the second derivative (acceleration) and first derivative (velocity) are included. However, the treatment of the white noise term remains fundamentally based on Gaussian random numbers, scaled appropriately to maintain consistency with the physical system being modeled.

\subsubsection{Simulation Interpretation}
The simulation results successfully represent the key characteristics of white noise and Brownian motion as described by Volpe and Volpe. In the white noise plots ( a–c), the generated noise is centered around zero, with the variance correctly scaling with 1/$\Delta t$. As the time step decreases, the noise becomes visibly more jagged, consistent with the theoretical expectation that a finer temporal resolution leads to rougher fluctuations.
\begin{figure}[h!]
    \centering
    \includegraphics[width=0.9\textwidth]{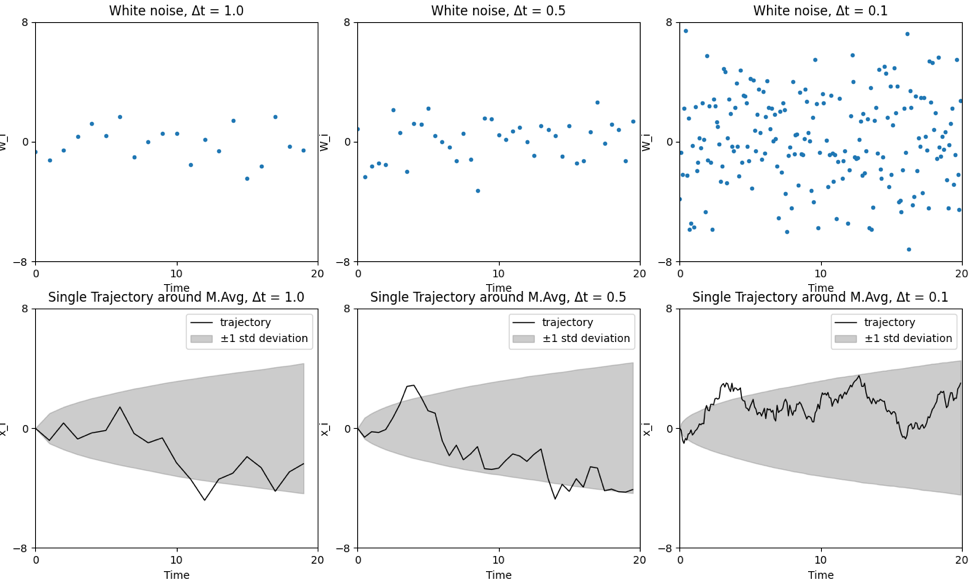}  
    \caption{Simulation of the White noise (position vs time).}
    \label{fig:trajectory}
\end{figure}
The corresponding random walk trajectories (d–f) exhibit stable statistical behavior across different $\Delta t$ values. Although the trajectories appear more irregular at smaller time steps, the overall variance remains consistent, indicating a fair representation of free diffusion. 

\subsection{Ballistic Motion to Brownian Motion}
Exploring the transition of a Brownian particle from ballistic motion to Brownian diffusion, closely following the theoretical framework and numerical approaches outlined in the paper. And the particle's behavior with and without considering inertial effects, leveraging stochastic forces modeled as white noise.
we numerically integrate the Langevin equation in two forms:
\subsubsection{With Inertia (Ballistic Motion):}
\[
m \ddot{x}(t) = -\gamma \dot{x}(t) + k x(t) + \sqrt{2 k_B T_\gamma} \, W(t)
\]
The deterministic terms (acceleration, friction, and restoring force) are approximated using finite-difference methods. The acceleration is replaced by:
\begin{equation}
\ddot{x}(t) = \frac{x_i - 2x_{i-1} + x_{i-2}}{(\Delta t)^2}
\label{eq:iterative_update}
\end{equation}
The velocity is approximated by Eq. (3). Thus, the discretized version of the Langevin equation describing the resulting motion without force term becomes: 
\begin{equation}
m \frac{x_i - 2x_{i-1} + x_{i-2}}{(\Delta t)^2} = -\gamma \frac{x_i - x_{i-1}}{\Delta t} + \sqrt{\frac{2 k_B T_\gamma}{\Delta t}} \, \eta_i
\label{eq:iterative_update}
\end{equation}
To scale the noise for discrete time steps, each random number is multiplied by:
\begin{equation}
\sqrt{ \frac{2 k_B T_\gamma}{m \left[1 + \Delta t \left(\frac{\gamma}{m}\right)\right]} \, \Delta t^{3/2} }
\label{eq:iterative_update}
\end{equation}
Thus, the solution for $x_i$ is:
\begin{equation}
x_i = \left( \frac{2 + \Delta t \left( \frac{\gamma}{m} \right)}{1 + \Delta t \left( \frac{\gamma}{m} \right)} \right) x_{i-1}
- \left( \frac{1}{1 + \Delta t \left( \frac{\gamma}{m} \right)} \right) x_{i-2}
+ \sqrt{ \frac{2 k_B T_\gamma}{m \left[ 1 + \Delta t \left( \frac{\gamma}{m} \right) \right]} \, \Delta t^{3/2} } \cdot w_i
\label{eq:iterative_update}
\end{equation}
\subsubsection{Without Inertia (Overdamped Diffusion):}
This is to ensures that the noise has the proper magnitude when integrated over time. In simpler random walk simulations (without inertia), the noise scaling simplifies to:
\begin{equation}
\dot{x}(t) = \sqrt{2D \, W(t)}
\label{eq:iterative_update}
\end{equation}
Where: D = is the diffusion coefficient $(D = \frac{k_B T}{c})$.\\
In this regime, inertial effects are negligible and the position is directly influenced by the diffusive term scaled by $\sqrt{2D}$.\\
The Eq. (10) becomes:
\begin{equation}
x_i = x_{i-1} + \sqrt{2 D \, \Delta t \, w_i}
\label{eq:iterative_update}
\end{equation}
Which is a good approximation for long time steps $(\Delta t \gg \tau)$, but it shows clear deviations at short time scale.\\
The mean square velocity autocorrelation function was defined as:
\begin{equation}
C_v(t) = \overline{v(t' + t) \, v(t')}
\label{eq:iterative_update}
\end{equation}
From the simulation $C_v$ becomes a discrete function, and decays to zero with the time constant represented in Fig. 2 (c).\\
The mean square displacement was defined as:
\begin{equation}
\langle x(t)^2 \rangle = \overline{ \left[ x(t' + t) - \overline{x(t')} \right]^2 }
\label{eq:iterative_update}
\end{equation}
And it can be calculated from a trajectory as:
\begin{equation}
\langle x_n^2 \rangle = \overline{ \left[ x_{i+n} - x_i \right]^2 }
\label{eq:iterative_update}
\end{equation}
Combining the deterministic and stochastic contributions, the position of the particle at each step is updated using Eq. (9). This relation allows the particle’s trajectory to be generated step by step, incorporating both the deterministic forces of the optical trap and the random kicks from thermal noise.\\
\begin{figure}[h!]
    \centering
    \includegraphics[width=0.9\textwidth]{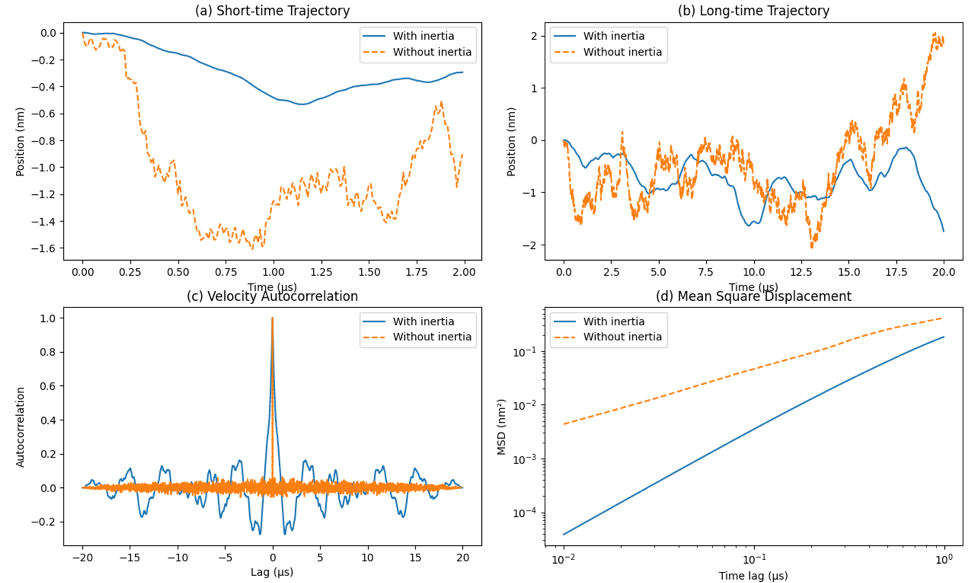}  
    \caption{}
    \label{fig:trajectory}
\end{figure}\\
Fig 2. (a) Shows short time trajectory, the inertial particle resists quick changes (due to mass), while the overdamped particle responds more sharply to random kicks. However, due to the zero initial velocity presented in this simulation, the overdamped motion starts slightly below.  (b) long-time trajectory Both lines (with and without inertia) show random walk behavior over longer times at long timescales; both systems become purely diffusive.  (c) velocity autocorrelation with inertia Shows a smooth, symmetric decay and without inertia Drops instantly to zero at zero lag. (d) mean square displacement with inertia: Initially quadratic growth (MSD $\sim t^2$), then transitions to linear (MSD $\sim t$). Without inertia: Linear growth (MSD $\sim t$) from the start. The inertial particle exhibits ballistic motion first, then diffusive motion and The overdamped particle is diffusive at all times.\\
The simulation captures the transition from ballistic to Brownian motion with strong agreement to theoretical expectations. At very short times, the inertial particle moves smoothly while the overdamped particle fluctuates more sharply, a behavior driven by their differing responses to thermal forces. A slight initial difference between the two trajectories, where the overdamped path starts lower, is likely due to idealized starting conditions and has no impact on the overall trends. Over longer times, both trajectories converge to random diffusion, losing any memory of initial velocity. The velocity autocorrelation and mean square displacement further confirm the shift from ballistic to diffusive behavior, matching the patterns described in the original study, the results validate the approach and faithfully reproduce the key physical features of the system.\\
\subsubsection{Optical Traps}
A Brownian particle trapped by optical tweezers is constantly buffeted by random thermal forces that push it away from the trap center, while optical forces work to restore it to equilibrium at the center. This interplay keeps the particle in a state of dynamic equilibrium.

The characteristic time scale for the restoring force to act is given by $\phi = \frac{\gamma}{k}$, where $\gamma$ is the friction coefficient and $k$ is the trap stiffness. This time scale $\phi$ is usually much longer than the particle’s momentum relaxation time $\tau = \frac{m}{\gamma}$, which is often negligible in typical experiments.

Hence, when modeling or simulating the dynamics of a Brownian particle in an optical trap, the particle’s position fluctuates due to thermal noise but is confined by the optical restoring force. The relevant time scale for the trap’s effect is $\phi = \frac{\gamma}{k}$, much larger than the inertial time $\tau$.

For accurate and stable numerical simulations, it is better to choose time steps such that $\tau \ll \Delta t \ll \phi$. Using a time step that is too large compared to $\phi$ leads to non-convergent, unphysical results.
\begin{figure}[h!]
    \centering
    \includegraphics[width=0.9\textwidth]{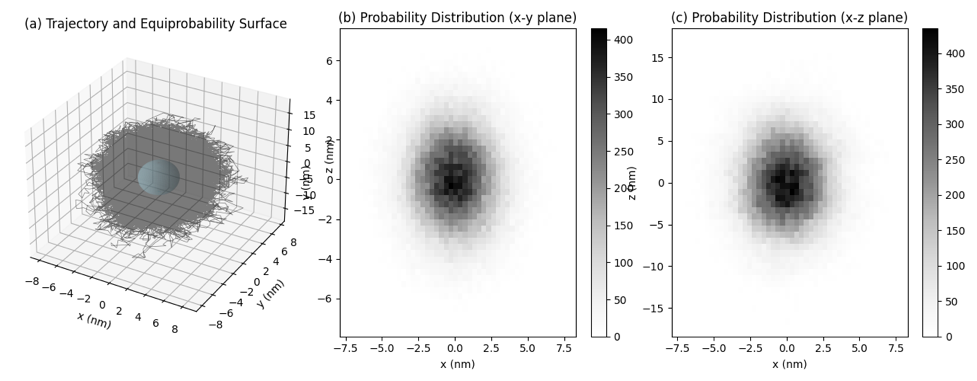}  
    \caption{(a) Trajectory and equiprobability surface. (b) and (c) probability of distribution}
    \label{fig:trajectory}
\end{figure}\\
The simulation successfully captures the behavior of a Brownian particle confined in an anisotropic optical trap. As shown in the trajectory plot, the particle randomly explores an ellipsoidal volume centered around the trap origin, with its motion bounded by the trap’s restoring forces.

The shaded equiprobability surface highlights the regions of highest likelihood, clearly showing an ellipsoidal shape that is wider along the $z$-axis, consistent with the weaker stiffness in that direction.

The probability distributions in the $x$–$y$ and $x$–$z$ planes further confirm this behavior, displaying Gaussian profiles centered at the origin, with broader spread along $z$ compared to $x$ or $y$.

These results accurately reproduce the physical features expected for a particle in a three-dimensional harmonic potential and closely match the experimental observations reported by Volpe and Volpe.
\subsection{Further Numerical Experiments:}
By extending the simulation approach for optically trapped Brownian particles, a wide range of more complex force scenarios can be explored beyond the simple harmonic trap. The general equation of motion can be written as:
\begin{equation}
\dot{x}(t) = \frac{1}{\gamma} F(x(t), T) + \sqrt{2D} \, W(t)
\label{eq:iterative_update}
\end{equation}
$F(x(t), t)$ is any force that may depend on both position and time. For a standard optical trap, this force is simply $-k x(t)$, but more elaborate situations can be modeled by modifying $F$.
\subsubsection{Constant Forces and Photonic Force Microscopy}
Introducing a constant force $F_c$ (applied at $t = 0$) shifts the equilibrium position of the trapped particle. The resulting displacement $\Delta x$ is directly proportional to the applied force via Hooke’s law: $F_c = k \Delta x$. 

This principle underpins photonic force microscopy, a technique widely used to measure tiny forces at the nanoscale, such as those exerted by bio-molecules. The effect is visible in the shift of the particle’s probability distribution within the trap.
\subsubsection{Non-Conservative and Rotational Forces}
In two dimensions, adding a rotational component to the force field-such as:
\begin{equation}
F(x, y) =
\begin{pmatrix}
k & -\gamma \Omega \\
\gamma \Omega & -k
\end{pmatrix}
\label{eq:force_matrix}
\end{equation}
With $\Omega$ representing the rotation rate, it causes the particle’s motion to exhibit rotational features. For strong rotation, the trajectory clearly shows circular motion, while for weak rotation, this is less apparent. The cross-correlation between $x$ and $y$ positions oscillates, reflecting the rotational dynamics \cite{Grier2003, Padgett2011, Risken1989}
\subsubsection{Double-Well Potentials and Kramers Transitions}
A double-well potential, created for example by two closely spaced optical traps, leads to two stable equilibrium positions separated by a barrier. The force in this case is:
\begin{equation}
F(x) = -a x^3 + b x
\label{eq:nonlinear_force}
\end{equation}
The particle can randomly hop between the wells-a phenomenon known as Kramers transitions. The frequency of these jumps depends on the barrier height and the temperature: lower barriers or higher temperatures increase the hopping rate. In a symmetric double-well, the average time spent in each well is the same, but adding a constant force can make the wells asymmetric, altering these residence times.
\subsubsection{Time-Dependent Potentials: Stochastic Resonant Damping and Resonance}
If the trap center itself oscillates in time, the force becomes:
\begin{equation}
F(x(t), t) = k \left[ x(t) - x_c \sin(2 \pi f t) \right]
\label{eq:time_dependent_force}
\end{equation}
When the oscillation frequency f matches the characteristic relaxation rate of the trap, the variance of the particle’s position can increase with trap stiffness-a counterintuitive effect known as stochastic resonant damping.\\
Stochastic resonance occurs when a particle in a double-well potential is subjected to a periodic driving force. The force then reads:
\begin{equation}
F(x(t), t) = -a x^3 + b x + c \sin(2 \pi f t)
\label{eq:general_force}
\end{equation}
If the driving frequency is close to the natural hopping rate between wells, the particle’s transitions can synchronize with the driving force. This synchronization is strongly temperature-dependent: too little thermal noise, and the particle rarely hops; too much, and the periodic force has little effect. There exists an optimal temperature where synchronization-and thus the system’s response to the periodic force-is maximized.
\section{Conclusion}
The representing of the simulation of a Brownian Particle in an Optical Trap successfully replicate the key phenomena described in the paper, capturing the dynamics of Brownian motion and behavior under an optical trap. The simulation of white noise accurately modeled thermal forces, while the transition from ballistic to diffusive motion aligned with expected theoretical trends. In the optical trap, the particle’s trajectories and probability distributions reflected the anisotropic stiffness and its effect on confinement, consistent with the paper's findings.

\section{Appedix}
\appendix
\section{Python Code: White Noise Simulation}
\begin{lstlisting}[style=pythonstyle]
import numpy as np
import matplotlib.pyplot as plt

# Set fixed random seed
np.random.seed(5000)

# Simulation parameters
timesteps = [1.0, 0.5, 0.1]  # Different Δt values
num_trajectories = 10000     # Number of trajectories for averaging

white_noises = []
trajectories = []
example_trajectories = []

# Simulate
for dt in timesteps:
    steps = int(20 / dt)  # Ensure x-axis covers 0 to 20
    wi = np.random.normal(loc=0.0, scale=np.sqrt(1/dt), size=steps)
    white_noises.append(wi)

    all_traj = np.zeros((num_trajectories, steps))
    for i in range(num_trajectories):
        w = np.random.normal(0, 1, steps)
        x = np.zeros(steps)
        for j in range(1, steps):
            x[j] = x[j-1] + np.sqrt(dt) * w[j]
        all_traj[i] = x
    trajectories.append(all_traj)
    example_trajectories.append(all_traj[0])  # Save one example trajectory

# Plotting
fig, axes = plt.subplots(2, 3, figsize=(18, 10))  # Create 2 rows x 3 columns of plots

# Subplots a, b, c: White noise (dots)
for i, ax in enumerate(axes[0]):
    x_vals = np.arange(0, 20, timesteps[i])
    ax.plot(x_vals, white_noises[i], 'o', markersize=3)
    ax.set_ylim(-8, 8)
    ax.set_xlim(0, 20)
    ax.set_xticks([0, 10, 20])
    ax.set_yticks([-8, 0, 8])
    ax.set_title(f"(a-c) White Noise, Δt = {timesteps[i]}")
ax.set_xlabel("Time")
    ax.set_ylabel("W_i")

# Subplots d, e, f: Single trajectory + shaded area ($\pm$ std deviation)
for i, ax in enumerate(axes[1]):
    x_vals = np.arange(0, 20, timesteps[i])
    mean_traj = np.mean(trajectories[i], axis=0)
    std_traj = np.std(trajectories[i], axis=0)
    example = example_trajectories[i]
    ax.plot(x_vals, example, 'k-', linewidth=1, label='Example trajectory')
    ax.fill_between(x_vals, mean_traj - std_traj, mean_traj + std_traj, color='gray', alpha=0.4, label='$\pm$1 std deviation')
    ax.set_ylim(-8, 8)
    ax.set_xlim(0, 20)
    ax.set_xticks([0, 10, 20])
    ax.set_yticks([-8, 0, 8])
    ax.set_title(f"(d-f) Single Trajectory + Shaded Area, Δt = {timesteps[i]}")
    ax.set_xlabel("Time")
    ax.set_ylabel("x_i")
    ax.legend()


plt.tight_layout()
plt.savefig('figure1_simulation.png', dpi=300, bbox_inches='tight')  plt.show()

\end{lstlisting}

\section{Python Code for Brownian Motion Simulation}

\begin{lstlisting}[language=Python, caption={}, basicstyle=\ttfamily\footnotesize, breaklines=true]
import numpy as np
import matplotlib.pyplot as plt

# Set fixed random seed
np.random.seed(5000)

# Constants
R = 1e-6  # meters
rho = 2200  # kg/m^3
m = 11e-15  # (4/3) * np.pi * R**3 * rho mass via density
g = 1e-3  # viscosity (Ns/m^2)
c = 6 * np.pi * g * R
T = 300  # Kelvin
kB = 1.38e-23
D = kB * T / c
tau = m / c

# Simulation settings
dt = 10e-9
N = int(20e-6 / dt)
time = np.arange(0, N) * dt

# Trajectory simulation
def simulate_trajectory(N, dt, with_inertia=True):
    x = np.zeros(N)
    if with_inertia:
        for i in range(2, N):
            w = np.random.normal(0, 1)
            x[i] = ((2 + dt * (c/m)) / (1 + dt*(c/m))) * x[i-1] \
                 - (1 / (1 + dt*(c/m))) * x[i-2] \
                 + (np.sqrt(2*kB*T*c)/m) * (dt**1.5) * w / (1 + dt*(c/m))
    else:
        for i in range(1, N):
            w = np.random.normal(0, 1)
            x[i] = x[i-1] + np.sqrt(2*D*dt) * w
    return x

# Full velocity autocorrelation
def full_velocity_autocorrelation(x, dt):
    v = np.diff(x) / dt
    v = v - np.mean(v)
    result = np.correlate(v, v, mode='full')
    result /= result[len(result)//2]
    lags = np.arange(-len(v)+1, len(v)) * dt
    return lags * 1e6, result  # lags in microseconds

# Mean square displacement
def mean_square_displacement(x, max_lag=100):
    msd = np.zeros(max_lag)
    for lag in range(1, max_lag):
        diffs = x[:-lag] - x[lag:]
        msd[lag] = np.mean(diffs**2)
    return msd

# Simulate trajectories
x_inertia = simulate_trajectory(N, dt, with_inertia=True)
x_no_inertia = simulate_trajectory(N, dt, with_inertia=False)

# Compute VACF and MSD
lags_inertia, vacf_inertia = full_velocity_autocorrelation(x_inertia, dt)
lags_no_inertia, vacf_no_inertia = full_velocity_autocorrelation(x_no_inertia, dt)
msd_inertia = mean_square_displacement(x_inertia)
msd_no_inertia = mean_square_displacement(x_no_inertia)
lags_msd = np.arange(len(msd_inertia)) * dt

# Plotting
fig, axs = plt.subplots(2, 2, figsize=(12, 10))

# (a) Short-time Trajectory
axs[0, 0].plot(time[:int(2e-6/dt)]*1e6, x_inertia[:int(2e-6/dt)] * 1e9, label='With inertia')
axs[0, 0].plot(time[:int(2e-6/dt)]*1e6, x_no_inertia[:int(2e-6/dt)] * 1e9, linestyle='--', label='Without inertia')
axs[0, 0].set_title('(a) Short-time Trajectory')
axs[0, 0].set_xlabel('Time ($\\mu$s)')
axs[0, 0].set_ylabel('Position (nm)')
axs[0, 0].legend()

# (b) Long-time Trajectory
axs[0, 1].plot(time * 1e6, x_inertia * 1e9, label='With inertia')
axs[0, 1].plot(time * 1e6, x_no_inertia * 1e9, linestyle='--', label='Without inertia')
axs[0, 1].set_title('(b) Long-time Trajectory')
axs[0, 1].set_xlabel('Time ($\\mu$s)')
axs[0, 1].set_ylabel('Position (nm)')
axs[0, 1].legend()

# (c) Full Velocity Autocorrelation
axs[1, 0].plot(lags_inertia, vacf_inertia, label='With inertia')
axs[1, 0].plot(lags_no_inertia, vacf_no_inertia, linestyle='--', label='Without inertia')
axs[1, 0].set_title('(c) Velocity Autocorrelation')
axs[1, 0].set_xlabel('Lag ($\\mu$s)')
axs[1, 0].set_ylabel('Autocorrelation')
axs[1, 0].legend()

# (d) Mean Square Displacement
axs[1, 1].loglog(lags_msd[1:] * 1e6, msd_inertia[1:] * 1e18, label='With inertia')
axs[1, 1].loglog(lags_msd[1:] * 1e6, msd_no_inertia[1:] * 1e18, linestyle='--', label='Without inertia')
axs[1, 1].set_title('(d) Mean Square Displacement')
axs[1, 1].set_xlabel('Time lag ($\\mu$s)')
axs[1, 1].set_ylabel('MSD (nm$^2$)')
axs[1, 1].legend()

plt.tight_layout()
plt.savefig('figure1_simulation.png', dpi=300, bbox_inches='tight')  
plt.show()
\end{lstlisting}

\section{Python Code: Optical Trap Simulation}
\begin{lstlisting}[style=pythonstyle]
import numpy as np
import matplotlib.pyplot as plt
from mpl_toolkits.mplot3d import Axes3D

# Set random seed for reproducibility
np.random.seed(5000)

# Physical constants
kB = 1.38e-23  # Boltzmann constant (J/K)
T = 300  # Temperature (K)
dt = 1e-6  # Time step (1 microsecond)
steps = 100000  # Number of steps
gamma = 1.88e-8  # Friction coefficient (kg/s)

# Trap stiffnesses (converted from fN/nm to N/m)
kxy = 1.0e-6 * 1e-6 / 1e-9
kz = 0.2e-6 * 1e-6 / 1e-9

# Precompute constants
sqrt_2kBT_gamma_dt = np.sqrt(2 * kB * T * gamma * dt)

# Initialize position arrays
x = np.zeros(steps)
y = np.zeros(steps)
z = np.zeros(steps)

# Langevin dynamics
for i in range(1, steps):
    wx, wy, wz = np.random.normal(0, 1, 3)
    fx = -kxy * x[i - 1]
    fy = -kxy * y[i - 1]
    fz = -kz * z[i - 1]

    x[i] = x[i - 1] + (fx / gamma) * dt + (sqrt_2kBT_gamma_dt / gamma) * wx
    y[i] = y[i - 1] + (fy / gamma) * dt + (sqrt_2kBT_gamma_dt / gamma) * wy
    z[i] = z[i - 1] + (fz / gamma) * dt + (sqrt_2kBT_gamma_dt / gamma) * wz

# standard deviations for ellipsoid
std_x = np.std(x)
std_y = np.std(y)
std_z = np.std(z)

# Create the ellipsoid
u = np.linspace(0, 2 * np.pi, 100)
v = np.linspace(0, np.pi, 100)
ellipsoid_x = std_x * np.outer(np.cos(u), np.sin(v))
ellipsoid_y = std_y * np.outer(np.sin(u), np.sin(v))
ellipsoid_z = std_z * np.outer(np.ones_like(u), np.cos(v))

# Plotting
fig = plt.figure(figsize=(18, 5))

# (a) 3D Trajectory and Equiprobability Ellipsoid
ax1 = fig.add_subplot(131, projection='3d')
ax1.plot(x * 1e9, y * 1e9, z * 1e9, color='k', alpha=0.5, lw=0.5)
ax1.plot_surface(ellipsoid_x * 1e9, ellipsoid_y * 1e9, ellipsoid_z * 1e9, color='lightblue', alpha=0.3,
                 edgecolor='none')
ax1.set_xlabel('x (nm)')
ax1.set_ylabel('y (nm)')
ax1.set_zlabel('z (nm)')
ax1.set_title('(a) Trajectory and Equiprobability Surface')

# (b) 2D Histogram: x-y plane
ax2 = fig.add_subplot(132)
hb1 = ax2.hist2d(x * 1e9, y * 1e9, bins=50, cmap='Greys')
ax2.set_xlabel('x (nm)')
ax2.set_ylabel('y (nm)')
ax2.set_title('(b) Probability Distribution (x-y plane)')
plt.colorbar(hb1[3], ax=ax2)

# (c) 2D Histogram: x-z plane
ax3 = fig.add_subplot(133)
hb2 = ax3.hist2d(x * 1e9, z * 1e9, bins=50, cmap='Greys')
ax3.set_xlabel('x (nm)')
ax3.set_ylabel('z (nm)')
ax3.set_title('(c) Probability Distribution (x-z plane)')
plt.colorbar(hb2[3], ax=ax3)

plt.tight_layout()
plt.savefig('figure1_simulation.png', dpi=300, bbox_inches='tight')  
plt.show()


\end{lstlisting}
\section*{Acknowledgments}
This work was originally developed as part of a research task during a PhD application process. Although it was not formally supervised, the author gratefully acknowledges the original study by Giorgio Volpe and Giovanni Volpe (2013), which served as the theoretical foundation for this computational reconstruction.
\clearpage
\bibliographystyle{apalike}  
\bibliography{reference}  

\end{document}